\documentclass[apj]{emulateapj}

\usepackage[letterpaper,portrait,left=1.25in,right=0.75in,top=1.5in,bottom=0.5in]{geometry}                
\usepackage{graphicx}
\usepackage{amssymb}
\usepackage{amsmath}
\usepackage{epstopdf}
\usepackage[para,online]{threeparttable}

\slugcomment{{\sc Accepted to AJ:} December 13, 2016}

\usepackage[backref,breaklinks,colorlinks,citecolor=blue]{hyperref}
\usepackage[all]{hypcap}

\usepackage{natbib}
\begin{document}
\title{C/O and O/H Ratios Suggest Some Hot Jupiters Originate Beyond the Snow Line}

%
\author{John M. Brewer, Debra A. Fischer}
	\affil{Department of Astronomy, Yale University}
	\affil{52 Hillhouse Avenue, New Haven, CT 06511, USA}
	\email{john.brewer@yale.edu}
	\email{debra.fischer@yale.edu}

\and

\author{Nikku Madhusudhan}
	\affil{Institute of Astronomy, University of Cambridge, Madingley Road, Cambridge CB3 0HA, UK}
	\email{nmadhu@ast.cam.ac.uk}

%


\begin{abstract}
The elemental compositions of planet hosting stars serve as proxies for the primordial compositions of the protoplanetary disks within which the planets form. The temperature profile of the disk governs the condensation fronts of various compounds, and although these chemically distinct regions migrate and mix during the disk lifetime, they can still leave an imprint on the compositions of the forming planets. Observable atmospheric compositions of hot Jupiters when compared against their host stars could potentially constrain their formation and migration processes. We compared the measured planetary and stellar abundances of carbon and oxygen for ten systems with hot Jupiters.  If the planets formed by core accretion with significant planetesimal accretion and migrated through the disk, the hot Jupiter atmospheres should be substantially super-stellar in O/H and sub-stellar in C/O.  On the contrary, however, we find that currently reported abundances of hot Jupiters have generally super-stellar C/O ratios, though present uncertainties on the reported O/H and C/O ratios are too large to reach a firm conclusion.  In one case however, HD~209458b, the elevated C/O and depleted O/H of the planet compared to the host star is significant enough to suggest an origin far beyond the ice line, with predominantly gas accretion, and subsequent disk-free migration. Improved measurements from the James Webb Space Telescope will enable more precise measurements for more hot Jupiters and we predict, based on the current marginal trend, that a sizable fraction of hot Jupiters will show enrichment of C/O and lower O/H than their hosts, similar to HD~209458b.
\end{abstract}


\maketitle

%
\section{Introduction} \label{sec:introduction}
Before the detection of 51 Peg b \citep{1995Natur.378..355M}, there was little evidence to support migration in planet formation models.  In the core accretion model described by \citet{1996Icar..124...62P}, the giant planets form outside the snow line and begin to accrete significant gas envelopes once the cores reach about 10 M$_{\oplus}$.  Terrestrial planet formation in this model can occur closer to the star because these planets don't accrete appreciable gas envelopes and so aren't constrained by gas disk lifetimes.  The discovery of a Jupiter mass planet on a 4 day orbit provided data to support theories of planet migration \citep{1996Natur.380..606L} and prompted the search for new ideas of how and where planets form.

As the number of planetary systems has increased, so have the questions about planet formation models.  The difficulty in growing dust beyond centimeter sized pebbles and the presence of tightly packed systems of rocky planets occupying the same orbital regions as hot Jupiter systems have prompted a re-examination of some underlying assumptions. Recently, there have been two new theories of in-situ formation for hot Jupiters \citep{2016ApJ...817L..17B,2015arXiv151109157B}.  These are still core accretion models, but the details of core formation change such that they no longer need to form outside the ice line.  One notable consequence of this change is where the planet accretes its final gas envelope.

During the evolution of the protoplanetary disk, the sequential condensation of solids results in changing abundance ratios for the remaining gas \citep{Moriarty:2014cb,2011ApJ...743L..16O}.  Thus, the location in the disk where a giant planet accretes its gaseous envelope can leave an imprint on its atmospheric composition.  Though the evolution of the disk combined with uncertainty in migration mechanism hinder locating a planet's exact birthplace \citep{2014ApJ...794L..12M}, certain regions of the disk are chemically distinct.  

Inside the water snowline, small amounts of oxygen will be bound up in magnesium silicates, but otherwise the gas disk should have a nearly stellar oxygen composition.  Between the water ice line at a few AU and the CO$_2$ ice line at $\sim 10$~AU, oxygen is bound up in water ice while carbon condensation is not energetically favored.  Thus, the gas in the region is oxygen poor, but carbon rich and the C/O ratio is significantly enhanced over its primordial and stellar abundance ratio  \citep{Piso:2015br,2011ApJ...743L..16O}.  Between the CO$_2$ and CO snow line at about 30~AU, twice as much oxygen as carbon condenses out, further decreasing the O/H ratio and enhancing the C/O ratio of the gas. Beyond the CO snow line, carbon and oxygen will be in solids reflecting the stellar abundance patterns and the gas will be depleted of any metals.

The location in the disk where the planet accretes the majority of its mass and the fraction of solid vs gas accretion will therefore influence the atmospheric composition of gas giant planets.  This natal imprint then gives us a clear discriminant in the case of hot Jupiter formation and subsequent disk-free migration from beyond the snow line.  If the C/O ratio of the hot Jupiter atmosphere is significantly elevated over that of its host star, then it was born beyond the water snow line. Gas giant planets with elevated C/O ratios and [O/H] less than the host star likely accreted their atmospheres beyond the water ice line and then experienced disk-free migration.  In contrast, gas giant planets with [O/H] comparable to or greater than their host star must have accumulated a significant fraction of solids or migrated early through the disk \citep{2014ApJ...794L..12M}.  

In the atmospheres of Jupiter and Saturn, the carbon abundance is enhanced by a factor of 3.5-7 over that in the solar photosphere \citep{2014prpl.conf..363P}. However, due to their cold temperatures, measuring their water (i.e. oxygen) abundance requires direct sampling \citep{Wong:2004dq}.  The situation is quite different for highly irradiated hot Jupiters where water has been clearly detected in 10 planets \citet{Kreidberg:2015er,Kreidberg:2014hi,Madhusudhan:2014ed,2014ApJ...783...70L}.

We compare the new stellar [O/H] and C/O abundances of \citet{Brewer:2016gf} to literature C/O and H$_2$O/H$_2$ values of hot Jupiter atmospheres and discuss current limitations and future prospects.

\begin{figure*}[htb!] 
   \centering
   \plotone{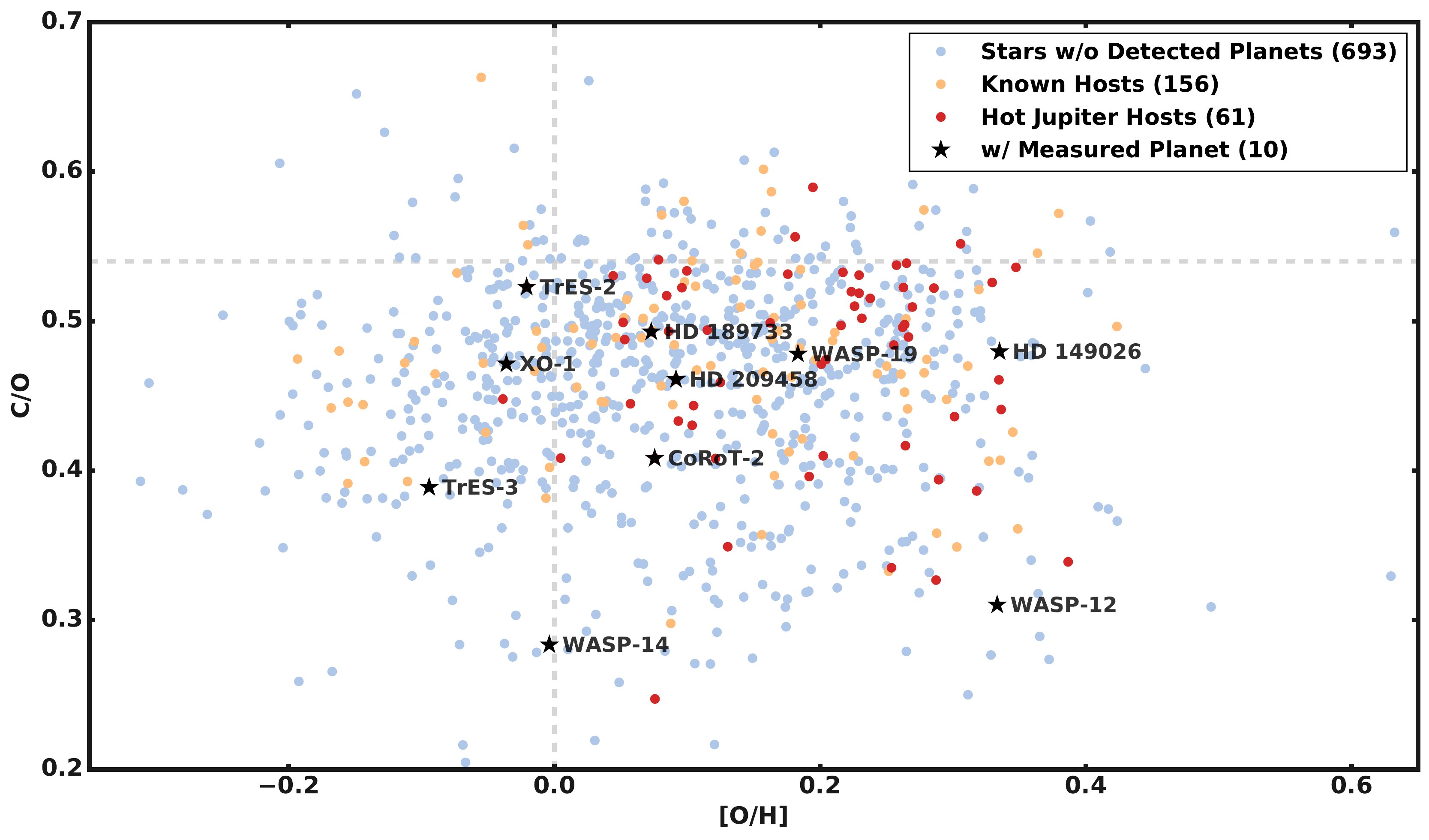} 
   \caption{C/O and [O/H] ratios of the 849 cool dwarfs from \citet{2016arXiv160806286B} and the hosts with measured planet atmosphere abundances used in this work.  The grey dashed lines mark the solar values of C/O and [O/H].  Stars without detected planets are designated by pale blue points, known planet hosts are in orange, and known hot Jupiter hosts are in red.  The known hot Jupiter hosts that also have measured C/O and [O/H] of the planet atmosphere are represented by a black star along with their names.}
   \label{fig:stellar_data} 
\end{figure*}

%
\section{Carbon and Oxygen} \label{sec:c_o_data}
The small number of unblended atomic lines of oxygen and carbon in optical spectra has made it challenging to derive accurate stellar abundances for C and O. For planets the task is further complicated by the high contrast ratios between planet and star, non-equilibrium chemistry, and clouds. In addition the hot Jupiter must transit in order to even obtain planetary atmospheric spectra.  Recent progress in measuring C and O for both stars and hot Jupiters is allowing us to move forward in our understanding of planet formation.

\subsection{Hot Jupiter Atmosphere Data} \label{sec:c_o_planets}
Secondary eclipse spectroscopy from a combination of ground and spaced based observatories gives us information about the day-side atmospheres of hot Jupiters.  The homogeneously analyzed catalog of \citet{2014ApJ...783...70L} provides C/O ratios and H$_2$O/H$_2$ for 9 exoplanet atmospheres and includes 1 $\sigma$ confidence regions.  An earlier study by \citet{Madhusudhan:2012ga} collected the results for six planet atmospheres, using a $\chi^2$ metric to distinguish between a carbon rich and an oxygen rich model for each planet.  One of the planets, HD 149026b, is smaller than the others at roughly a Saturn-mass though the arguments for its atmospheric composition are the same and so we include it and refer to it as a hot Jupiter for the sake of completeness.  

In addition to the eclipse measurements, transmission spectroscopy is available for several planets \citep{Kreidberg:2015er,Kreidberg:2014hi,Madhusudhan:2014ed}.  These measurements give us information about the atmosphere at the day-night terminator of the planet.  The planets studied overlap with those observed with eclipse spectroscopy and two of the planet atmospheres were analyzed using both eclipse and transit data to derive self consistent atmospheric models \citep{Kreidberg:2015er,Kreidberg:2014hi}.

Seven of the planets have atmospheric H$_2$O/H$_2$ measurements.  Since N(H) $\gg $ N(O), the hydrogen in the water will be negligible and we can estimate the O/H ratio of the planets as $\sim 0.5 \times$ H$_2$O/H$_2$.  A comparison of the planet O/H ratio to that of its host star provides a constraint on how the C/O ratio was enhanced or depleted.  This in turn could potentially inform us about the formation conditions of the planetary atmosphere \citep{2011ApJ...743L..16O}.

\subsection{Host Star Data} \label{sec:c_o_hosts}
The stellar properties catalog of \citet{Brewer:2016gf} contains abundances for 15 elements including carbon and oxygen.  The authors note that corrected trends exist over some subsets of the temperature ranges and that these regions should only be used with caution in comparative analyses.  For [C/H] and [O/H] the data are relatively free of trends below 6100K.  This affects both WASP-12 at 6154~K and WASP-14 at 6459~K, though we have chosen to keep WASP-12 in this study as it lies just at the edge of the acceptable range.  Nine stars remain in common between the stellar host measurements and the hosts of those with planet atmosphere measurements: HD 189733, HD 209458, HD149026, WASP-12, WASP-19, TrES-2, TrES-3, XO-1, and CoRoT-2.  \citet{Brewer:2016gf} derived empirical errors for each of their elements, with [C/H] having an error of 0.026 dex and [O/H] an error of 0.036 dex, resulting in a typical error in C/O of $\sim$~10\% and [O/H] of $\sim$~8\%.

%
\section{Results} \label{sec:results}

\subsection{Implications of Stellar Abundances for hot Jupiter Compositions}
\label{sec:results_stellar}

\begin{figure*}[htb!] 
   \centering
   \plotone{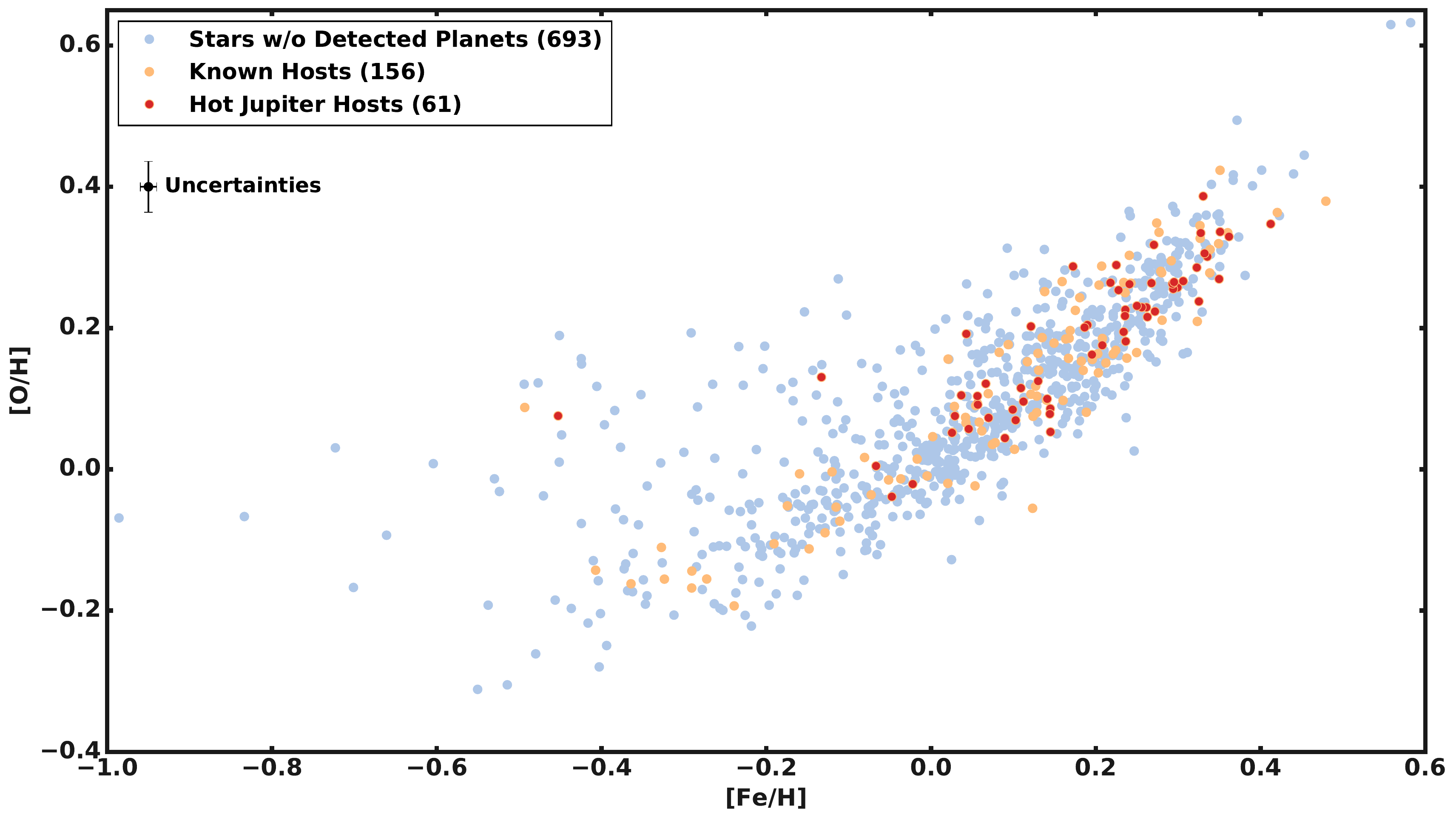} 
   \caption{Stellar oxygen vs. iron abundance for the 849 cool dwarfs from \citet{2016arXiv160806286B} and the hosts with measured planet atmosphere abundances used in this work.  Colors are the same as in Figure \ref{fig:stellar_data}. In general, the oxygen abundance is tightly correlated with the metallicity as measured by [Fe/H].  Since hot Jupiters are preferentially found around metal rich stars \citep{2005ApJ...622.1102F}, their hosts will also be oxygen rich.}
   \label{fig:o_vs_fe_data} 
\end{figure*}

The C/O and [O/H] ratios of the planet hosting stars in our sample are shown in Fig.~\ref{fig:stellar_data}, based on the stellar abundance data of \citet{Brewer:2016gf} and the sub-sample of cool dwarfs with well measured carbon and oxygen in \citet{2016arXiv160806286B}. We find two important trends in the distribution of elemental abundances of planet hosting stars which have important consequences for exoplanetary compositions. Firstly, the distribution of C/O ratios across our sample shows that the sun is moderately carbon-rich. Most stars, including planet-hosting stars, have lower C/O ratios (i.e. are more oxygen-rich) compared to the sun, i.e. C/O $<$ 0.54. In particular, the ten stars hosting hot Jupiters considered in this work have sub-solar C/O ratios. Secondly, we also find that most of the planet-hosting stars, including most of the hosts of the hot Jupiters in this study, also have super-solar O/H ratios. This follows the previously known general trend that the occurrence rate of giant exoplanets increases with host stellar metallicity, [Fe/H] \citep{2005ApJ...622.1102F}, and hence with the O/H ratio, since we find that [O/H] scales with [Fe/H] (Figure \ref{fig:o_vs_fe_data}). These trends suggest that the formation environments of giant exoplanets might be predominantly oxygen-rich. 

\subsection{Oxygen Rich Core Accretion}
\label{sec:oxygen_rich_accretion}

The sub-solar C/O ratios and super-solar oxygen abundance of most planet hosting stars lead to observable consequences for hot Jupiter compositions.  Under the standard model of giant planet formation with core accretion, the metallicity of the planetary atmosphere is driven largely by the composition of the solids accreted in planetesimals during formation and migration through the disk.  Giant planets forming in such an environment could then be  expected to be even more enriched in oxygen, i.e. [O/H] $\gg 0$ and C/O $< 0.54$. The resultant atmosphere is expected to be enhanced in all the condensible elements relative to the host star \citep{2009EGUGA..11.6324A, 2012ApJ...751L...7M, 2009AA...507.1671M, 1999Natur.402..269O, 1996Icar..124...62P}, as has been observed for C, N, S, P, and inert gases in Jupiter's atmosphere \citep{2016arXiv160604510A,1999Natur.402..269O}. Since oxygen is the largest constituent of planetesimals, primarily via H$_{2}$O ice and silicates, the atmosphere is particularly enriched in oxygen. Here it is assumed that the elemental composition of the nebula and hence the protoplanetary disk mimic the stellar composition. 

The oxygen-rich elemental compositions thus predicted for hot Jupiters imply that H$_2$O is the most abundant volatile in their atmospheres, and is also the most observable molecule with current instruments \citep{2014ApJ...783...70L,2016SSRv..tmp...31M}. It is well known that in H$_2$-dominated atmospheres with O-rich compositions (e.g. C/O~$\lesssim 0.8$)  H$_2$O is the most dominant O-bearing molecule in the observable atmosphere at all temperatures ($\sim$~300--3500~K). Thus for hot Jupiters with super-solar O/H and sub-solar C/O the H$_2$O abundance should be super-solar, with volume mixing ratios H$_2$O/H$_2$~$\gtrsim$~10$^{-3}$ \citep{Madhusudhan:2012ga}; at high temperatures ($\gtrsim$1200 K) a fraction of the O ($\lesssim$50\%) can also be bound in CO. Such abundances of H$_2$O are easily observable in transmission and emission spectra of hot Jupiters.

If, however, hot Jupiters are found to be under-abundant in H$_2$O that would imply different formation and migration conditions than those assumed in the standard picture discussed above. In general, it would indicate a paucity of planetesimals accreted during the formation and/or disk-free migration, e.g. via dynamical encounters \citep{2014ApJ...794L..12M}. In what follows, we compare constraints on atmospheric abundances of several hot Jupiters to those of their host stars to constrain the different formation and migration pathways. 

\begin{figure*}[htb!] 
   \centering
   \plotone{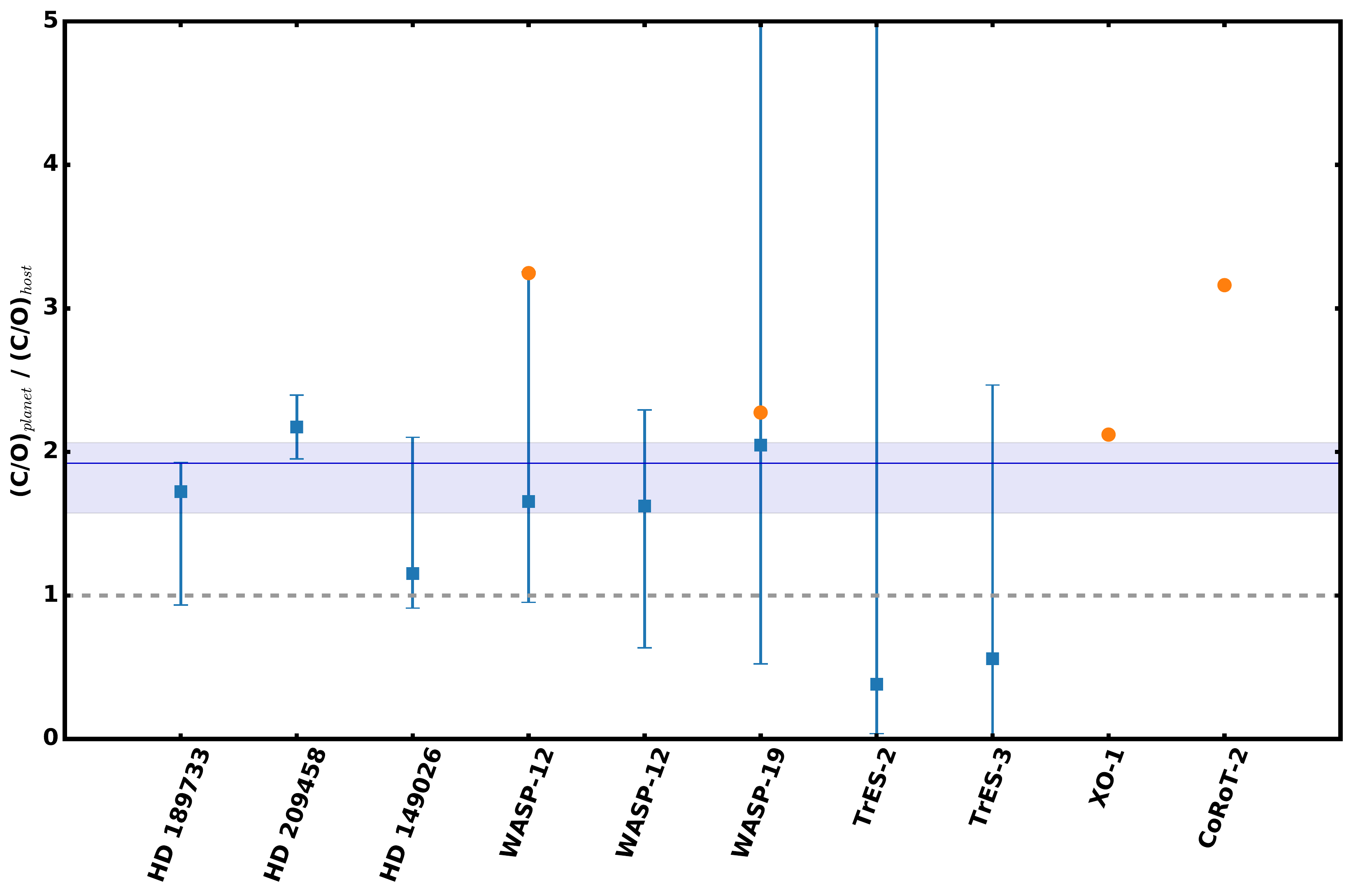} 
   \caption{Ratios of hot Jupiter C/O to those of their stellar host.  The blue squares with error bars use the hot Jupiter data from \citet{2014ApJ...783...70L} and the orange circles are best fit values from \citet{Madhusudhan:2012ga} and do not have published uncertainties.  The upper range of the uncertainties for WASP-19 and TrES-2 are 10.5 and 15.4 respectively and the plot range has been limited for clarity.  The grey dashed line marks the 1:1 ratio where planet and host have the same C/O ratio. The blue solid line and shaded region is a fit to the mean for the blue points and their uncertainties.  The average planet C/O ratio is super-stellar, though the large uncertainties make all but one consistent with their host star.}
   \label{fig:co_ratio_compare}
\end{figure*}

\subsection{Planetary Vs Stellar Abundances}
\label{planetary_vs_stellar}

In this section we investigate how C/O ratios of hot Jupiters compare with C/O ratios of their host stars. Typically, constraints on atmospheric properties of transiting hot Jupiters are obtained using transmission spectra of the day-night terminator, when the planet transits in front of the host star, and/or thermal emission spectra from the dayside atmosphere, when the planet is at secondary eclipse. Given an observed spectrum, the atmospheric chemical compositions and 1-D averaged temperature profile are derived using atmospheric retrieval methods (see e.g. reviews by \citet{2014prpl.conf..739M,2016SSRv..tmp...31M}). In planetary atmospheres the dominant elements are all in molecular form, due to which the elemental abundances are derived from the respective major molecular species. In particular, the C/O ratios are derived from abundances of H$_2$O, CO, CH$_4$, and, to some extent, CO$_2$, which are the  most prominent O and C bearing species in H$_2$-rich atmospheres under a wide range of atmospheric conditions \citep{2016SSRv..tmp...31M,Madhusudhan:2012ga,2013ApJ...763...25M}. 

Reliable constraints on C/O ratios of hot Jupiters are possible only when multi-band atmospheric observations are available at wavelengths which contain strong spectral features of H$_2$O, CO, and CH$_4$. Currently, such data, and hence constraints on the C/O ratios, are available for several hot Jupiters primarily from thermal emission photometry and/or spectra of their dayside atmospheres obtained using the Spitzer and Hubble space telescopes. In the present study we use the C/O ratios of such systems reported in recent literature \citep{2014ApJ...783...70L,Madhusudhan:2012ga}. On the other hand, several recent studies have also reported precise atmospheric observations of transmission spectra but such data only contain constraints on the H$_2$O abundances. In principle, the H$_2$O abundances can be used to derive constraints on the C/O ratios under certain assumptions of chemical and radiative equilibrium (see e.g. \citet{2015arXiv150407655B}), but we do not use those C/O ratios in the present study in the interest of being free from model assumptions. 

We compared stellar C/O ratios obtained using the stellar abundance data of \citet{Brewer:2016gf} with the hot Jupiter C/O ratios from the atmospheric retrieval methods of \citet{2014ApJ...783...70L,Madhusudhan:2012ga}. Fig.~\ref{fig:co_ratio_compare} shows the planetary vs stellar C/O ratios for 9 systems, though two only have best fit measurements for planetary C/O with no published uncertainties. The most precisely measured exoplanet atmosphere is that of HD 209458b, which has a C/O ratio 2.2 times that of its host star.  All but two of the remaining also have super-stellar C/O ratios.  The large uncertainties in the planetary atmosphere measurements mean that all planets except HD 209458b are consistent with having the same C/O ratios as their hosts. Using only the measurements with reported uncertainties in planet C/O we calculated the mean planet-to-star ratio as 1.9 $^{+0.14}_{-0.35}$, indicating that the planetary C/O ratios tend to be higher than their host stars. This suggests that while the stars themselves are oxygen-rich, i.e. have low C/O ratios as discussed in section~\ref{sec:results_stellar}, the planets are preferentially oxygen-poor or carbon-rich.

\begin{figure*}[htb!] 
   \centering
   \plotone{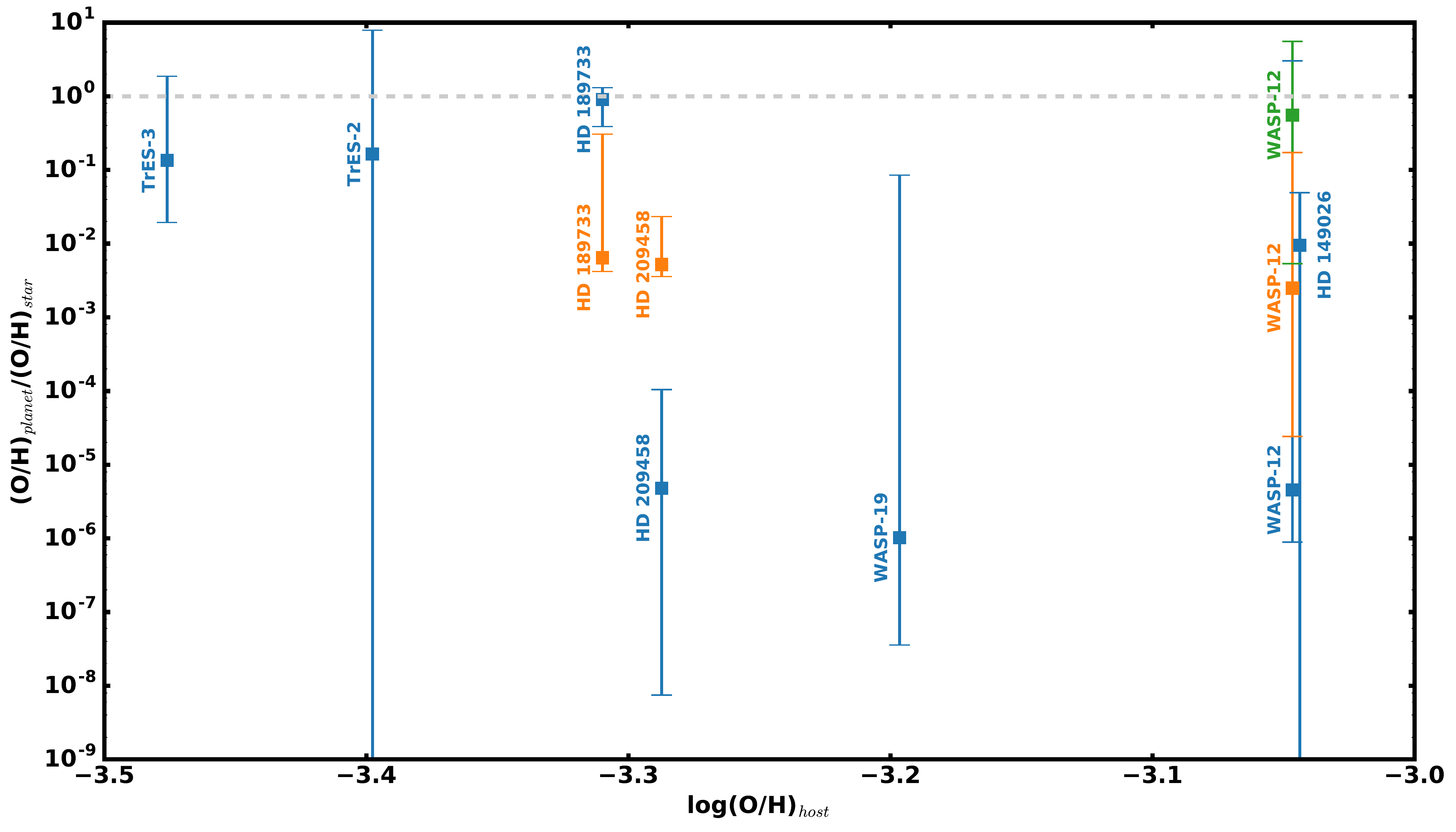} 
   \caption{Ratios of hot Jupiter O/H values with those of their host stars.  The planet O/H ratios are estimated from their H$_2$O/H$_2$ measurements.  Several of the planets were analyzed by multiple groups, in blue are those from \citet{2014ApJ...783...70L}, orange points are from \citet{Madhusudhan:2014ed}, and green from \citep{Kreidberg:2015er}.  For reference, we have placed a grey dashed line indicating the 1:1 ratio between planet and host star, though clouds, gravitational settling, and the precise volume mixing ratios in the planet atmosphere will alter the H$_2$O abundance seen in the planet atmospheres for a given initial composition.}
   \label{fig:oh_ratio_compare}
\end{figure*}

We also investigate trends in the O/H ratios of hot Jupiters versus their parent stars. To this end we use H$_2$O measurements reported for 7 hot Jupiters, using emission and/or transmission spectra, from which we derive the O/H ratio assuming H$_2$O is the most dominant oxygen carrier (\S \ref{sec:oxygen_rich_accretion}). Figure~\ref{fig:oh_ratio_compare} shows the comparisons of the stellar vs planetary O/H ratios. Four of the seven planets have O/H ratios consistent with their host stars. The \citet{2014ApJ...783...70L} measurement of HD~189733b shows a marginally super-stellar O/H ratio, though the \citet{Madhusudhan:2012ga} measurement from transmission spectra is almost 3 orders of magnitude smaller but still consistent with its host. Both HD~149026b and WASP-19b have lower oxygen abundances than their host stars, though the measurements also have very large uncertainties. HD 209458b shows an unambiguous detection of sub-stellar O/H with both transmission and emission spectroscopy. Combined with its super-stellar C/O ratio, this planet is consistent with formation beyond the snow line and subsequent disk-free migration \citep{2014ApJ...794L..12M}.

Our results above are critically reliant on the accuracy of the derived molecular abundances in hot Jupiter atmospheres. While our preliminary analysis above suggests that hot Jupiters are generally oxygen-poor relative to their host stars, more  detailed observations are required to improve their abundance constraints. Additionally, recent studies have suggested the possibility of clouds/hazes in some hot Jupiters which could in principle influence the abundance determinations, particularly  from transmission spectra where clouds/hazes are expected to have the most impact \citep{2016SSRv..tmp...31M,2015MNRAS.446.2428S}. For example, in some cases the low H$_2$O abundances derived from transmission spectra in the near-infrared using the HST WFC3 instrument could potentially be due to clouds/hazes partially obscuring the atmospheres \citep{Madhusudhan:2014ed}. However, this is less likely to be the case in thermal emission spectra which probe the very hot dayside atmosphere of the planet and for which the observations are in the mid-infrared where clouds/hazes are expected to have the least impact. Overall, however, the fact that our conclusions are consistent between both the transmission and emission abundance estimates suggest that hot Jupiters might indeed be oxygen-poor relative to their host stars. 

%
\section{Discussion} \label{sec:discussion}

Elemental abundance ratios of the host stars provide a proxy for the primordial abundances in the protoplanetary disk and, hence, for the initial chemical conditions of planetary formation. Several recent studies have investigated the range of outcomes of giant planetary compositions  depending on their formation locations and migration paths \citep{2011ApJ...743L..16O, Helling:2014hb, 2014ApJ...794L..12M,Piso:2015br,2016arXiv160903019M}. It is generally assumed that the atmospheric composition reflects the accretion history of the planet \citep{Pinhas:2016ka}. 

The key parameters determining the atmospheric composition of a hot Jupiter are the location and amounts of solids accreted relative to the gas in forming the giant planet.  For a solar composition disk, with a C/O ratio of 0.5, oxygen-rich species dominate the solid composition \citep{2011ApJ...727...77M,2012ApJ...751L...7M,2011ApJ...743..191M,2012ApJ...757..192J,2014A&A...570A..35M,2015A&A...574A.138T}. Within the H$_2$O snow line ($\lesssim$2 AU), the gas is expected to be of nearly stellar composition since the only condensates are silicates which contain $\sim$20\% of the O. Further out in the disk the gas and solid compositions are governed by the locations relative to the major snow lines; the gas becomes progressively C-rich reaching a C/O ratio of $\sim$1 beyond the CO$_2$ snow line, and the solids become progressively O-rich reaching a stellar C/O ratio beyond the CO snow line where nearly all volatiles are condensed out. Therefore, hot Jupiters forming within the H$_2$O snow line are expected to be of nearly stellar-composition. On the other hand, hot Jupiters forming beyond the H$_2$O snow line with significant planetesimal accretion are always expected to be O-rich, with a maximum C/O that of the nebula when formed beyond the CO snow line.

The planet-host stars in our sample are even more oxygen-rich, with higher [O/H] and lower C/O ratios, compared to the sun. In the standard core accretion scenario, the corresponding hot Jupiters can be expected to be substantially super-solar in oxygen, and hence in H$_2$O abundances.  Such oxygen enrichment in hot Jupiters is readily detectable with current and future instruments.  Therefore, if hot Jupiters are found to have sub-stellar oxygen, it would indicate low planetesimal accretion, formation beyond the H$_2$O snow line, and a migration path that inhibits solid accretion, e.g. by dynamical scattering mechanisms rather than migration through the disk \citep{2014ApJ...794L..12M}.  

The precision of most hot Jupiter atmosphere C/O measurements is not currently high enough to rule out in situ formation models or migration through the disk.  However, the precise measurements of HD 209458b suggests that at least some hot Jupiters form beyond the snow line and migrate without accreting significant oxygen rich solids.  As the measurement precision improves, it is our prediction that most if not all planet atmosphere measurements will show super-stellar C/O ratios and/or sub-stellar O/H ratios.  The James Webb Space Telescope (JWST) will be able to obtain much more detailed spectra, with $100 <= \lambda/\delta\lambda < 3600$ between 1 and 11$\mu$ \citep{Greene:2016gx}, of hot Jupiter atmospheres and allow us to test this hypothesis.

As discussed above, the amount of enhancement can also tell us about the migration history of the planets.  If migration begins before the dissipation of the gas disk, then the hot Jupiter will accrete gas over a wide range of semi-major axes.  This will dilute the signal, perhaps erasing it altogether.  The fact that we see a clear signal in HD 209458b tells us that at least in some cases, the majority of the planet atmosphere is accreted beyond the snow line without much pollution from oxygen rich planetesimals.  Additional hot Jupiters with measured H$_2$O abundances were published \citep{2016arXiv161001841B} during the review process of this paper.  They also show a tendency toward sub-solar O/H ratios, suggesting disk-free migration.

\subsection{Solar Versus Stellar}
Most previous studies of hot Jupiter atmospheres have assumed solar abundance patterns and asked whether the C/O ratio was Sun-like (0.5) or super-solar (e.g. $\sim$1). Although these questions are interesting, using the \textit{stellar} abundance pattern and looking for deviations in the planet atmosphere with respect to its own star can tell us a lot more about the planet formation process in these systems  \citep{2014ApJ...788...39T}.  Noting that the C/O ratio of WASP-12b is approximately solar tells us little about how it may have formed, though does tell us that any H$_2$O in its atmosphere should be readily detectible. Realizing that the planet C/O may be almost twice its stellar C/O ratio can help identify where in the disk it formed.

\section{Conclusions} \label{sec:conclusions}

The hot Jupiter, HD 209458b, has a C/O ratio 2.2 times that of its host star and a sub-stellar O/H ratio.  Though all other planets in this study are consistent with having the same C/O ratios as their host star, the mean ratio of planet-to-star shows an enhancement in C/O.  This would be a clear signal of disk-free migration from beyond the snow line and rule out in situ formation of hot Jupiters for most stars.  Improved precision in measuring the composition of exoplanet atmospheres with instruments such as JWST will be able to resolve this issue.  Stellar hosts of hot Jupiters are preferentially oxygen rich and standard core accretion models would enhance the oxygen abundance in the planet atmosphere.  A deficit of oxygen would signal low levels of planetesimal accretion or disk-free migration. C/O ratios in planet atmospheres show a larger range than that of host stars.  This may be telling us something fundamental about the formation processes of the planets.

%
\acknowledgments
This work was supported by a NASA Keck PI Data Award, administered by the NASA Exoplanet Science Institute. Data presented herein were obtained at the W. M. Keck Observatory from telescope time allocated to the National Aeronautics and Space Administration through the agency's scientific partnership with the California Institute of Technology and the University of California. The Observatory was made possible by the generous financial support of the W. M. Keck Foundation. This research has made use of the SIMBAD database, operated at CDS, Strasbourg, France 

The authors wish to recognize and acknowledge the very significant cultural role and reverence that the summit of Mauna Kea has always had within the indigenous Hawaiian community. We are most fortunate to have the opportunity to conduct observations from this mountain.

JMB and DAF thank NASA grant NNX12AC01G.

%
\bibliography{ms}

\typeout{get arXiv to do 4 passes: Label(s) may have changed. Rerun}
\end{document}